\documentstyle[aps,prl,epsf,rotate]{revtex}
\begin{document}

\twocolumn[\hsize\textwidth\columnwidth\hsize\csname     
@twocolumnfalse\endcsname

\title{Reorientation of Anisotropy in 
a Square Well Quantum Hall Sample 
}

\author{W. Pan$^{1,2}$,
T. Jungwirth$^{3,4}$, H.L. Stormer$^{5,6}$,
D.C. Tsui$^{1}$,
A. H. MacDonald$^{3}$,
S. M. Girvin$^{3}$,
L. Smr\v{c}ka$^{4}$,
L.N. Pfeiffer$^{5}$, K.W. Baldwin$^{5}$, and
K.W. West$^{5}$}
\address{$^{1}$Department of Electrical Engineering, Princeton
University,
Princeton, New Jersey 08544}
\address{$^{2}$NHMFL, Tallahassee, Florida 32310}
\address{
$^{3}$Department of Physics,
Indiana University, Bloomington, Indiana 47405}
\address{$^{4}$Institute of Physics ASCR,
Cukrovarnick\'a 10, 162 00 Praha 6, Czech Republic}
\address{$^{5}$Bell Labs, Lucent Technologies, Murray Hill, New
Jersey 07974}
\address{$^{6}$Department of Physics
and Department of Applied Physics, Columbia University, New
York,
New York 10027}

\date{\today}
\maketitle

\begin{abstract}

We have measured magnetotransport at half-filled high Landau
levels in a
quantum well with two occupied electric subbands. We find
resistivities that are {\em isotropic} in perpendicular
magnetic field but become strongly {\em anisotropic}
at $\nu$ = 9/2 and 11/2 on tilting the field.
The anisotropy appears at an in-plane field, $B_{ip} \sim 2.5$
T, 
with the easy-current direction {\em parallel} to $B_{ip}$ but
rotates by 90$^{\circ}$ at $B_{ip} \sim 10$ T and points now in
the same direction as in single-subband samples.
This complex behavior is in quantitative agreement with
theoretical
calculations based on a unidirectional charge density wave
state model.

\end{abstract}

\vskip2pc]

A two-dimensional (2D) electron gas is an attractive system for
many-body physics studies\cite{qhe:book,fqhe:book}. 
A particularly 
rich variety of phenomena associated with strong interactions
among
electrons
appears in the regime of the fractional quantum
Hall effect (FQHE) \cite{tsui:prl82,laughlin:prl83}. 
During much of the last decade, studies of the
FQHE have focused on
even-denominator Landau level filling factors
\cite{sarma:book96,heinonen:book98} such as  
the compressible
$\nu=1/2$ state and the $\nu=5/2$  incompressible
quantum Hall fluid.
Most recently, strongly
anisotropic transport has been observed 
in high
quality GaAs/Al$_x$Ga$_{1-x}$As single heterojunctions
\cite{horst:aps93,lilly:prl99a,du:ssc99,pan:prl99,lilly:prl99b}
at  filling factors
$\nu$ = 9/2, 11/2, etc., and in 2D hole
systems\cite{shayegan:physicae99} 
starting at $\nu=5/2$.
In these experiments, the magnetoresistance shows a
strong peak in one current direction and a deep minimum in the
perpendicular current direction.  Tilting
the magnetic field away from the sample normal causes the
high resistance direction to change from its original
orientation  
to the in-plane magnetic field direction. 

The origin of the magnetotransport anisotropy has not been
firmly 
established yet. The most appealing interpretation suggests 
that the 2D electron gas
spontaneously breaks the translational symmetry by forming
a unidirectional charge density wave (UCDW), as predicted by
Hartree-Fock theory
\cite{koulakov:prl96,moessner:prb96}.  
This idea has spurred much theoretical interest 
\cite{fradkin:prb99,fertig:prl99,rezayi:prl99,simon:preprint99,jungwirth:prb99,philip:prl00,fradkin:preprint99,rezayi:preprint99,macdonald:preprint99,maeda:preprint99,yu:preprint99,vonoppen:preprint99,wang:preprint99,cote:preprint00}.
Because of uncertainty about the reliability of this Hartree-Fock
prediction,
there has been a special emphasis 
\cite{jungwirth:prb99,philip:prl00} 
placed on tests of its ability to explain 
experimental results on ``stripe'' orientation in tilted
magnetic fields. 
In particular, Jungwirth et al.
\cite{jungwirth:prb99} carried out
detailed many-body RPA/Hartree-Fock calculations combined with a
self-consistent
local-spin-density-approximation (LSDA) description of
one-particle
states in experimental sample geometries. For the sample
parameters of the 
traditional, single-interface specimens of Refs. 
\cite{pan:prl99,lilly:prl99b} with a single electric subband
occupied, the theory \cite{jungwirth:prb99,philip:prl00}
gives stripes oriented 
perpendicular to the field, consistent with experiment.

A theoretical study \cite{jungwirth:prb99} of UCDWs 
in  parabolic quantum wells 
that have two subbands occupied in zero magnetic field, has
predicted
much more complex behavior of the UCDW state, including stripe
states induced
by an in-plane field and rotation of stripe orientation at
critical
in-plane field strengths.  A comparison between theory and
experiment in a geometry for which this intricate behavior
occurs, 
constitutes an excellent test of the
UCDW explanation of anisotropic transport in higher Landau
levels.
Since parabolic quantum wells are experimentally difficult to
realize and suffer from poor mobility we, instead, chose a
square
well structure which is expected to exhibit similarly complex
behavior, provided that more than one electric subband is
occupied in zero
field.

Our sample, detailed in Figure~1(c), consists of a 350\AA~wide
GaAs quantum well bracketed between
thick Al$_{0.24}$Ga$_{0.76}$As layers grown on a (100) GaAs
substrate by MBE. Two Si delta-doping layers are placed
symmetrically above and below the quantum well at a distance of
800\AA. The specimen has a size of 5mm $\times$ 5mm and is
contacted via eight indium contacts, placed symmetrically around
the perimeter.
The electron density is established after illuminating the
sample
with a red light-emitting diode at $\sim$4.2K and we measure an
electron mobility of $\mu$ = 7 $\times$ 10$^6$ cm$^2$/V~s. The
total electron density, $n = 4.6 \times 10^{11}$ cm$^{-2}$, is
determined from low-field Hall
data. The subband densities, $n_1 = 3.3\times 10^{11}$ cm$^{-2}$
and $n_2=1.3 \times
10^{11}$ /cm$^{-2}$, are obtained 
by Fourier analysis of the low-field Shubnikov-de Haas
oscillation. Their values coincide with the results of our
numerical self-consistent LSDA calculation. 
All angular-dependent measurements were carried out 
at T = 40 mK in a top-loading dilution
refrigerator equipped with an {\it in-situ} rotator
\cite{eric:rsi99}
placed inside a 33T resistive
magnet.
A low-frequency ($\sim$ 7 Hz) lock-in
technique at a current $I$ = 10 nA is used.
We define the axis of rotation as the {\it y}-axis.
Consequently, the in-plane field, $B_{ip}$, is along
the {\it x}-axis when the sample is rotated. 
Therefore, $R_{xx}$ refers to ``$I$ parallel to
$B_{ip}$'' and $R_{yy}$ refers to ``$I$ 
perpendicular to $B_{ip}$'' \cite{remm}. 

Figure~1(a) shows an overview of magnetoresistance at zero-tilt.
The shaded region highlights the transport features around $\nu$
=
9/2 and 11/2. The integer quantum Hall effect (IQHE) states at
$\nu
= 1,2,3, \cdots$ and the FQHE states at $\nu$ = 2/3, etc. are
clearly visible. Figure~1(b) shows the results of
self-consistent
LSDA calculations of Landau levels (measured from the bottom of
the
quantum well) in perpendicular magnetic field.

Figure~2 shows the $R_{xx}$ and $R_{yy}$ data 
for $4 < \nu < 6$ at four different
tilt angles $\theta$ = 0$^{\circ}$, 41.2$^{\circ}$,
67.9$^{\circ}$,
and 76.2$^{\circ}$. The tilt angle is determined using the shift
of
prominent QHE states, which depend only on the
perpendicular magnetic field, $B_{perp} = B~\times$ cos$\theta$.

In the absence of $B_{ip}$ ($\theta$ =
0$^{\circ}$), $R_{xx}$ and $R_{yy}$ show
a peak at $\nu$ = 9/2 and a slight dip at $\nu$ = 11/2 and
negligible anisotropy. The small
difference in magnitude between $R_{xx}$ and $R_{yy}$ 
is probably a result of the different contacts involved in both
measurements.
This practically {\em isotropic} behavior of $R_{xx}$ and
$R_{yy}$ 
is distinctively different from 
results
\cite{horst:aps93,lilly:prl99a,du:ssc99,pan:prl99,lilly:prl99b}
on single-subband, single-heterojunctions, where the
states at $\nu$ = 9/2 and 11/2 are
strongly anisotropic in the absence of $B_{ip}$. This lack of 
anisotropy in our sample has a simple interpretation. The
diagram in Figure~1(b) 
indicates that the $\nu$ = 9/2 and 11/2 states 
are the $\nu$ = 3/2 state of
the lowest Landau level (N=0) in the second quantum well 
subband (i=2). The $\nu$ = 3/2 state in single subband samples
exhibits isotropic transport, which seems to
carry over to the second subband. Yet, exceptional behavior
develops 
on tilting the specimen.

At $\theta=41.2^{\circ}$ the $R_{xx}$ and $R_{yy}$ traces
are very different from  
those taken at zero field-tilt and different from each other.
The $\nu$ = 9/2 and $\nu$ = 11/2 states are strongly
anisotropic with the hard-axis perpendicular to $B_{ip}$
($R_{yy}$)
and the easy-axis parallel to $B_{ip}$ ($R_{xx}$).
The direction of this tilt-induced anisotropy (TIA) 
is {\em rotated by 90$^{\circ}$} as
compared to the direction in traditional single-subband,
single-heterojunction structures 
\cite{pan:prl99,lilly:prl99b}.
As the tilt angle increases further, the $R_{xx}$ and $R_{yy}$
traces approach each other again at $\theta \sim 67.9^{\circ}$
rendering the transport nearly isotropic (Figure~2(c)). 
Beyond this angle the anisotropy reemerges but 
the hard-axis and easy-axis {\em have traded places}, as seen in
Figure~2(d). 

In Figures~3(a,b) we plot $R_{xx}$ and
$R_{yy}$ at filling factors $\nu$ = 9/2 and 11/2 versus
$B_{ip}$. Their general behavior is rather
similar. Practically isotropic transport prevails in the range
of $0 < B_{ip} <
2$~T, but there is a clear onset to anisotropy at $B_{ip} \sim
2.5$~T. The level of anisotropy rapidly increases, reaching its
peak at $B_{ip} \sim 5.0$~T, whereupon
the $R_{xx}$ and $R_{yy}$ values approach each other again and
cross at $B_{ip} \sim 10$~T. For higher in-plane fields the
transport is 
again anisotropic, but its direction has {\em rotated by
90$^{\circ}$}.
Figures~3(c,d) show
the anisotropy factor, defined as
$(R_{xx}-R_{yy})/(R_{xx}+R_{yy})$
and derived from the data of the panels above. They clearly
depict
the initially, practically isotropic behavior followed by a strong
anisotropy that
rotates direction by 90$^{\circ}$ at $B_{ip} \sim 10$~T.
The direction of anisotropy in single-subband samples corresponds
to the high $B_{ip}$ direction in our double-subband specimen.

We now turn to the analysis of correspondence between
the measured TIA and the theory
based on the UCDW picture. 
For an infinitely narrow electron layer the effective 2D Coulomb
interaction, $V(\vec{q})$, reduces to
$e^{-q^2\ell^2/2}/q\,(L_N(q^2/2))^2\,
2\pi e^2\ell/\epsilon$ where $L_N(x)$ is the
Laguerre polynomial, $\vec{q}$ is the wavevector, $\ell$ is the
magnetic length,
and $\epsilon$ is the dielectric function. 
Starting from $N=1$, zeros of $L_N(q^2/2)$ occur at
finite $q=q^*$, producing a zero in the repulsive Hartree
interaction at  wave vectors where the attractive
exchange interaction is strong. For the half-filled valence
Landau level the
corresponding UCDW state
consists of alternating
occupied and empty stripes of electron guiding center states
with
a modulation period
$\approx 2\pi/q^*$.

In finite-thickness 2D systems subjected to
tilted magnetic fields, the dependence of the effective
interaction on wavevector magnitude $q$ and orientation $\phi$
relative to the in-plane field direction
can be accurately approximated \cite{jungwirth:prb99} by
$V(\vec{q})=V_0(q)+
V_2(q)\cos(2 \phi)$. At $B_{ip}=0$,
the isotropic term $V_0(q)$
has a wavevector-dependence similar to that of the effective
interaction in
the infinitely narrow 2D layer.
The corresponding curve for the valence Landau level at
$\nu=9/2$, shown in the top inset of
Figure~4,
has no zeros at finite $q$-vectors
{\em because} the half-filled valence Landau level is the $N=0$
state of the
second subband (as shown in detail in Figure~1(b)) \cite{rem}.
Hence, the UCDW state is not expected to form, consistent
with the isotropic transport measured in perpendicular field.

Because of the finite thickness of the 2D system in our
350\AA~wide quantum well,
the orbital effect of the in-plane field causes Landau levels
emanating from different
electric subbands to coincide, depending on the strength of
$B_{ip}$.
The in-plane field
mixes electric and magnetic levels so the subband and
orbit radius indices are no longer good quantum numbers.
However,
the effect of $B_{ip}$ near the level (anti)crossing can
sometimes 
be viewed approximately as a transfer of valence electrons from
the
lowest ($N=0$) Landau level of the second subband
to a higher ($N>0$)  Landau level of the first subband.
For filling factor $\nu=9/2$,  such a circumstance occurs in our
sample   
at $B_{ip}\approx3$ T, as seen from the top and bottom insets of
Figure~4.
Indeed, $V_0(q)$ is only slightly modified at low in-plane
fields,
while a clear minimum 
develops for $B_{ip}>3$ T. As discussed above for the case of
perpendicular
magnetic field it is the minimum of the interaction energy at
finite wavevector
that opens the possibility for the formation
of the UCDW state. The theoretical and experimental critical
in-plane fields
corresponding to the onset of the UCDW and TIA, respectively,
are remarkably close.

The non-zero anisotropy coefficient $V_2(q)$ of the effective
interaction at $B_{ip}>0$
is responsible for the
non-zero UCDW anisotropy energy $E_A$, defined
\cite{jungwirth:prb99} as the
total Hartree-Fock energy of stripes oriented parallel with
$B_{ip}$
minus the total energy of stripes  perpendicular to $B_{ip}$.
The direction of the anisotropy results from a delicate 
competition between electrostatic and exchange contributions to
$E_A$ and 
can be determined only by an accurate calculation which takes
into
account  details of the experimental configuration.
As shown in Figure~4, the stripes align
parallel with $B_{ip}$ at low in-plane fields, consistent with
the
measured
easy-current direction parallel with $B_{ip}$. The sign of the
UCDW
anisotropy
energy changes at $B_{ip}=10$ T which coincides with the
experimental critical field
for the interchange of easy and hard current axes.
This theoretical discussion of the $\nu=9/2$ state was found
to apply for $\nu=11/2$ as well. 

In conclusion, we have observed complex transport behavior
in a two-subband QW at half-filled high Landau levels.
Both the transition to an anisotropic transport state,
at finite $B_{ip}$, and the rotation of the direction 
of anisotropy by 90$^{\circ}$ at higher $B_{ip}$ are explained
quantitatively by the UCDW picture.
The close agreement between
complex experimental data and theoretical results leaves little
doubt as to the origin of the observed transport anisotropies
in high Landau levels.

We would like to thank E. Palm and T. Murphy for experimental
assistance, and N. Bonesteel, R.R. Du, and K. Yang for useful discussion.
A portion of this work was performed at the National
High Magnetic Field Laboratory which is supported by NSF
Cooperative
Agreement No. DMR-9527035 and by the State of Florida.
The work at Indiana University was supported by NSF grant
DMR-9714055,
and at the Institute of Physics ASCR
by the Grant Agency of the Czech Republic
under grant 202/98/0085.
D.C.T. and W.P. are supported by the DOE and the NSF.

\vspace*{-0cm}

\begin{figure}[t]
\epsfxsize=3.0in
\centerline{
\epsffile{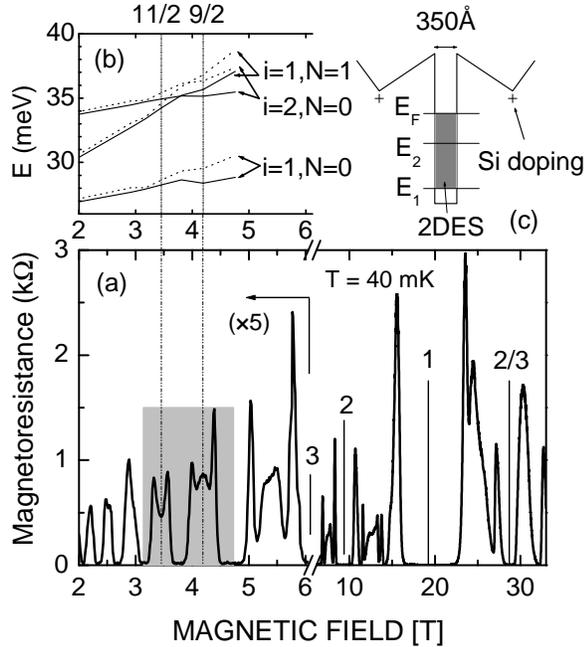}}

\vspace*{-0cm}

\caption{
(a) Overview of magnetoresistance in perpendicular magnetic
field. The IQHE states ($\nu$ = 1, 2, 3, etc.) and the FQHE
states
($\nu$ =
2/3, etc.) are marked by vertical lines. Shaded region
highlights the transport features around $\nu$ = 9/2 and 11/2.
(b) Self-consistent LSDA 
energy levels in perpendicular field.
Index of electric subband (i) and Landau level (N) is shown for
each energy level. Solid lines represent the spin-up state and
dotted lines represent the spin-down state. (c)
Structure of our quantum well sample. The well width is
350$\AA$. E$_{F}$, E$_2$, and E$_1$ are the zero-field
Fermi energy, second, and
first subband energy level, respectively. 
}
\end{figure}

\vspace*{-0.3cm}

\begin{figure}[t]
\epsfxsize=3.0in
\centerline{
\epsffile{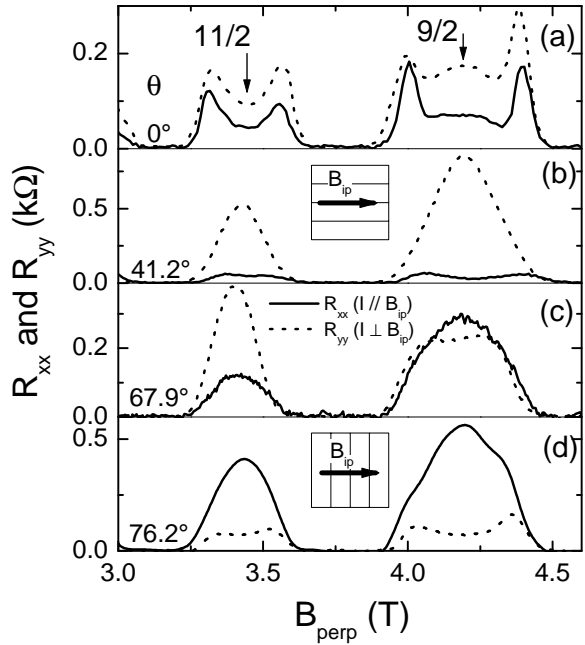}}

\vspace*{-0.0cm}

\caption{
Magnetoresistance $R_{xx}$ (solid lines) and $R_{yy}$ (dotted
lines)
between $4 < \nu < 6$ as a function of perpendicular magnetic
field,
$B_{perp}$, at four tilt angles, $\theta$.
The in-plane magnetic field $B_{ip}$ is along the {\it
x}-axis. Stripes in the insets of panels (b) and (d) indicate
the
tilt-induced anisotropy (TIA) at $\nu$ = 9/2 and 11/2.
}
\end{figure}

\vspace*{-0cm}

\begin{figure}[t]
\epsfxsize=3.0in
\hspace*{-0cm}
\centerline{
\epsffile{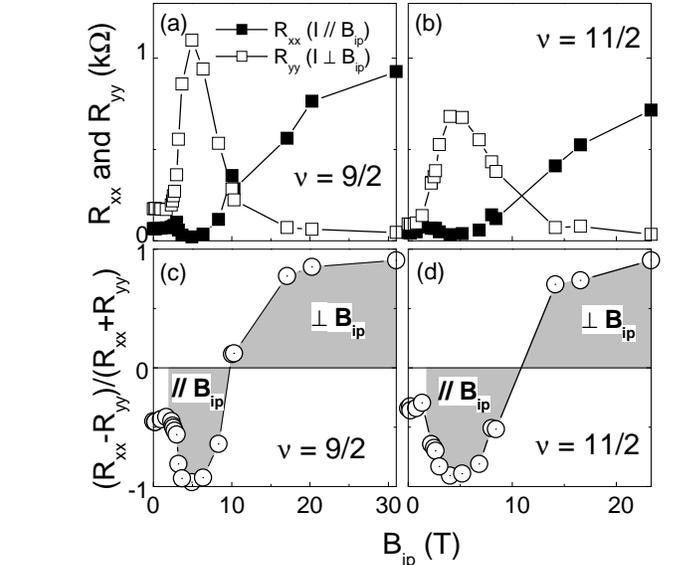}}

\vspace*{-0cm}

\caption{
Amplitude of $R_{xx}$ (solid squares) and $R_{yy}$
(open squares) at $\nu$ = 9/2 [panel (a)] and
at $\nu$ = 11/2 [panel (b)] as a function of $B_{ip}$. Panels
(c)
and (d) show the anisotropy factor, defined by 
$(R_{xx}-R_{yy})/(R_{xx}+R_{yy})$ and derived from the data of
the
panels above.
The shade regions represent the tilt-induced anisotropy (TIA)
parallel
with and perpendicular to $B_{ip}$, respectively.}
\end{figure}

\vspace*{-0.5cm}

\begin{figure}[t]
\epsfxsize=3.3in
\hspace*{-0.5cm}
\centerline{
\epsffile{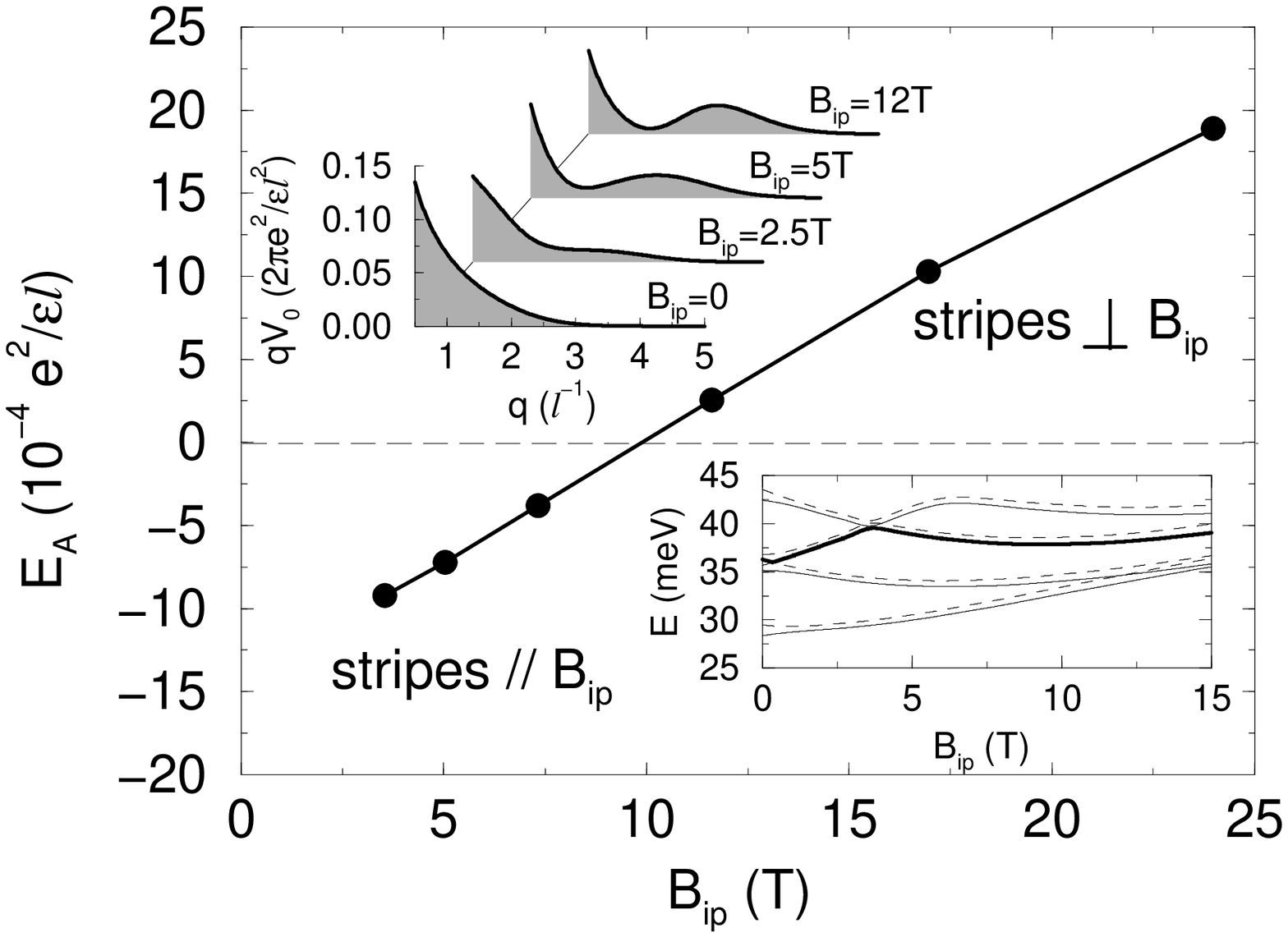}}


\caption{Theoretical results for the $\nu=9/2$ state. Main
graph:
UCDW anisotropy energy as a function of in-plane magnetic
field.
Top inset: 
the isotropic term of the effective
2D Coulomb interaction multiplied by the wavevector amplitude
$q$ at
different in-plane fields. Bottom inset: self-consistent LSDA
Landau
levels as a function of in-plane magnetic field. Thick line is
the half-filled
valence Landau level.
}
\end{figure}


\begin{references}
\bibitem{rem} Landau levels in perpendicular and tilted magnetic
fields 
were calculated using the full self-consistent
local-spin-density-approximation
(LSDA) method. Effective 2D Coulomb interaction curves
were obtained from the LSDA  one-particle 
orbitals. The screening
from electrons occupying the lower filled Landau levels was
accounted
for using the RPA. For details of the calculations see  
\protect\cite{jungwirth:prb99}.
\end{references}
\end{document}